\documentclass[doublecol]{epl2}
\pdfoutput=1

\newcommand\poi{Poincar$\acute{\rm e}$~}

\def\ket#1{| \,#1\, \rangle}

\def\expect#1{\langle \,#1\, \rangle}

\newcommand\hbtd{{\rm{HB-T}}}
\newcommand\hbt{{\rm{HB-T}~}}

\title {{Classical light analogue of the nonlocal 
Aharonov-Bohm effect}}
\shorttitle{{Classical light analogue of the nonlocal 
Aharonov-Bohm effect}} 

\author{Nandan Satapathy$^{1}$,  Deepak Pandey$^{1}$ \and Poonam
Mehta$^{1,2}$  \and
Supurna Sinha$^{1}$ \and \\
Joseph Samuel$^{1}$ \and Hema Ramachandran$^{1}$}
\shortauthor{N. Satapathy \etal}

\institute{\inst{1}
  Raman Research Institute, C. V. Raman Avenue, Sadashivanagar, Bangalore, India 
560 080.\\
  \inst{2} Presently at the Department of Physics and Astrophysics,
University of Delhi, Delhi, India 110 007. }


\abstract{
 We demonstrate the existence of a non-local geometric phase in  the
intensity-intensity correlations of classical incoherent light, that is
not seen in the lower order correlations. This two-photon 
Pancharatnam
phase was observed and modulated in a Mach-Zehnder interferometer.  Using acousto-optic interaction, independent 
phase noise was introduced to light in the two arms of the interferometer to create two independent incoherent classical
sources from laser light. The experiment is the classical optical
analogue of the multi-particle Aharonov-Bohm effect.  As the trajectory
of light over the \poi sphere introduces a phase shift observable
only in the intensity-intensity correlation, it provides a means of
deflecting  the two-photon wavefront, while having no effect on single photons.} 

\begin{document}

\maketitle

{\it Introduction:} 
Two classic interference experiments are Young's 
double-slit experiment and the Hanbury Brown and Twiss (\hbtd) 
experiment.
The former measures the amplitude-amplitude correlation 
and demonstrates the interference of a photon with itself.  The 
latter experiment measures the intensity-intensity
correlation and is a 
demonstration of the interference of a pair of
photons with itself. While the \hbt intensity-interferometry experiment was
historically important in leading to our present understanding of  
quantum optics and coherence, 
one does not need to invoke quantum mechanics to understand it; 
it can quite simply be understood entirely 
in terms of classical electric field fluctuations.  
In recent times, with the development of different types of 
light sources and detection techniques, intensity-interferometry has, once again, become a topic of great interest. It has led to  many
tantalizing ideas and interesting applications \cite{Strekalov, Saleh, Pittmann, Bouwmeester, shihbook}.  
There has been enormous debate \cite{Bennick-Boyd, Abouraddy,Chapter-21,unified-theory} on the similarities and differences between bi-photon interferometry and two-photon inerferometery. Bi-photon interferometry \cite{Klyshko, Burlakov, Shih-review} uses entangled photon pairs, {\it i.e.,} pairs of photons that are related due to a conservation principle, for example, product photons in parametric down-conversion. Two-photon interferometry \cite{EPL_Scarcelli-Shih}, on the other hand,  involves  photons that are statistically correlated, as from a thermal source; our present experiment belongs to this category. Intensity-interferometry with classical light is regaining importance with the demonstration of ghost imaging with thermal light \cite{Valencia, Friberg, Zhou, Karmakar, Dheera}.   

\begin{figure}
\onefigure[scale=0.3]{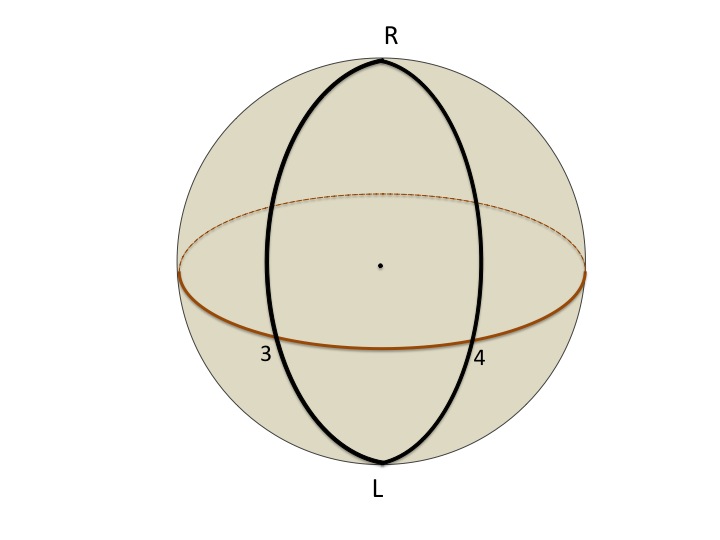}
\caption{(Colour online) The \poi sphere. Figure shows four points on the \poi sphere
representing two circular polarisation states ($R,  L$) and two linear 
polarisation states of the two polarisers in front of the two detectors  
($3, 4$). The geometric phase 
is given by half the solid angle subtended at the centre, by the geodesic  polygon  
R-4-L-3-R.} \label{fig.poi}
\end{figure}

A classical \hbt experiment  consists of  two slits,  each illuminated by an independent, 
incoherent light source. Two detectors measure the intensity of light falling on them; at each 
detector, the intensity has contributions  from both slits.   
The cross correlation of the intensities at the
two detectors varies in a sinusoidal fashion as a function of separation between the 
detectors. In the conventional \hbt experiment, the appearance of these
fringes was a purely dynamical effect, arising from the change in the 
path difference of the two detectors from the two slits.
Recently it was shown theoretically \cite{Mehta}
that even with fixed detectors, one can have such fringes 
if the  sources and detectors are polarised and the polarisation is
varied.
This is purely a  geometric or Pancharatnam phase effect~\cite{panch,punch} 
and  arises due to the closed trajectory of polarised light 
on the \poi sphere. While such geometric phase effects are well known
in amplitude interferometry, their counterparts in intensity 
interferometry have been less studied.  
The geometric phase is given by
half the solid angle subtended by a closed path traced out by polarising 
elements on the \poi sphere (see Fig.~\ref{fig.poi}). 
Unlike the dynamical phase shift that is restricted by the
spatial and temporal  coherence of the sources, the geometric phase, 
is achromatic and   unbounded, as  will be discussed shortly.

 In this Letter, we wish to convey several points. We demonstrate 
a new way of producing classical light with exactly tailored 
characteristics.  This is achieved by creating radio-frequency electrical signals of the desired features and imprinting this onto light via acousto-optic interaction.    {We use this technique to create light that has random phase noise only. Next, }we show that a small modification to the \hbt intensity 
interferometry experiment leads to an optical analogue of the multiparticle 
Aharonov-Bohm (AB) effect \cite{samuelsson}. 
This manifests as a non-local geometric phase in  the
intensity-intensity correlations of classical incoherent light, that is
not seen in the lower order correlations. Finally, we show 
that this non-local two-photon cross-correlation 
can be modulated via the geometric phase.  This suggests a 
way of deflecting the  cross-correlated photon pairs.

 {The AB effect \cite{ab, SamCS} shows that the electromagnetic potential potential affects charged particles even in regions where no field exists. It is seen as a phase acquired by the wave function of an electron when  it moves in a path enclosing  a solenoid. The path is entirely in a region where the  magnetic field is zero and the acquired phase is attributed to the enclosed flux \cite{PT}. The AB effect  led to a wide 
appreciation of the role of the electromagnetic potential,
in particular its interpretation \cite{wuyang} as a 
``connection'',  a 
structure well studied by mathematicians in differential geometry. 
A connection is a rule for comparing quantities defined at different
points\footnote {A connection need not be integrable along 
a closed curve and this lack of integrability is called a holonomy. If 
the connection is locally integrable it has no curvature and is said to 
be flat.}.  The AB effect can be viewed 
as a physical manifestation of a flat connection, 
since there is no magnetic field in the region accessible to
the electron.  The AB effect has both geometric and topological
aspects to it.
It is insensitive to small deformations of the interfering
paths and therefore may be viewed as topological.  At the same time, it {\it is} sensitive to small changes of the magnetic flux and in this sense is 
geometric.

 Essentially the same mathematical structure underlies the Pancharatnam phase, seen, for example\cite{Sam-Hema}, in a Sagnac interferometer, where the dynamic path difference is zero.  This geometric phase is a consequence of the solid angle enclosed on the \poi sphere.  It too is insensitive to small changes in the paths of the interfering beams and hence  is {\it topological}, while being  sensitive to small changes in the solid angle enclosed on the \poi sphere  is also {\it geometric}. 
There is thus a close analogy between the Pancharatnam phase and the AB
phase.\\ }
 {While the Pancharatnam phase predates the AB phase,  the latter has recently been 
extended to {\it non local} multiparticle effects which intriguingly, are manifest {\it only} in the cross correlations, and not in the 
self correlations or in the lower order correlations. Considered 
theoretically for a pair of electrons by Samuelsson {\it et al} \cite{samuelsson},  this two-particle non-local AB effect  was observed experimentally
by Neder {\it et al}\cite{heiblum}.  A pair of electrons  together enclosed an 
AB flux due to an applied magnetic field, which was  then used to modulate the 
two particle cross correlation via coupling to the  orbital degree of freedom of electrons while the spin remained frozen.  The effects of the flux were absent in the individual currents and in their self correlations while the cross correlation between electron currents revealed the dependence on the
AB flux. }

The experiment of Neder {\it et al}  was 
essentially quantum in nature,  as electrons, being fermions, do not admit the limit of large particle numbers. In contrast to fermions, bosons permit 
macroscopic occupancy of a single state, admitting a 
classical limit  and permitting a  classical field 
theory description.  The experiment that we report here  is the classical bosonic analogue of the two-particle non-local
AB effect.  { Classical light beams from two sources meet after having traversed two independent paths over the \poi sphere and together enclose a solid angle on the sphere. One of the polarising elements  was varied so as to modulate the enclosed solid angle, and the first order correlation (interference visibility),  the second order self-correlation  and the second order cross-correlations were measured. The effect of the modulation of the geometric phase was seen only in the second order cross correlation and not in the other measured quantities. Thus,   apart from  the sign differences owing to 
the different statistics of the particles involved, the 
effect that we report here and the two-particle non-local AB effect  are conceptually very similar. }

\begin{figure}
\onefigure[scale=0.3]{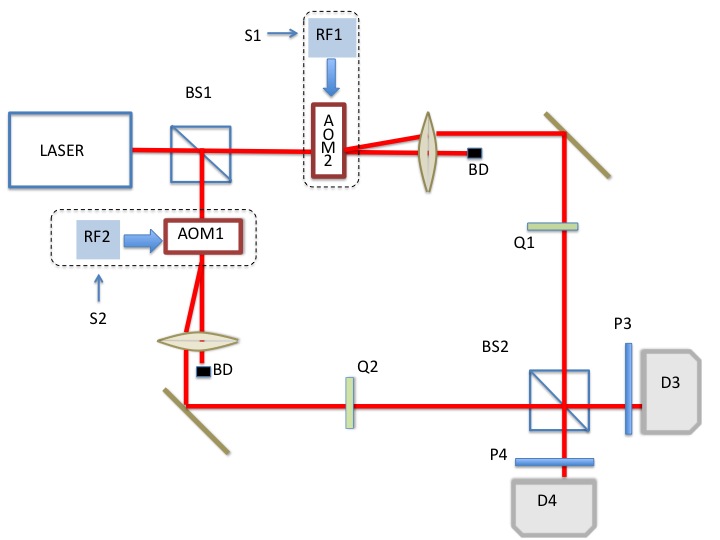}
\caption{(Colour online) Figure shows a schematic diagram of the
experimental setup. Two sources $S1$ and $S2$ illuminate two 
detectors $D3$ and $D4$ as explained in the text.  {$BS1$ and 
$BS2$ are 
non-polarising beam splitters, $Q1$ and $Q2$  are quarter wave 
plates,  
$P3$ and $P4$  are linear polarisers and $BD$  are beam dumps.}} 
\label{fig2.schematic}
\end{figure}

{\it Experimental Setup:}
Our experiment employs two phase-incoherent  sources $S1$ and $S2$   
in a Mach-Zehnder setup with  birefringent elements in the  two arms, and polarisers preceding two detectors 
 $D3$ and $D4$ (see Fig.~\ref{fig2.schematic}). 
A variation of the relative polarisation angle of the detectors 
introduces a geometric phase equal to half the
solid angle enclosed by the two interfering paths
on the \poi sphere (see Fig.~\ref{fig.poi}). 
Light  from a diode laser at 767nm  was incident on a non-polarising 
beam-splitter  depicted as BS1 in Fig.~\ref{fig2.schematic},
and the two emergent beams were passed through two 
acousto-optic-modulators (AOM) 
shown in Fig.~\ref{fig2.schematic} as AOM1 and AOM2, 
fed by two radio frequency sources 
(RF1 and RF2) respectively. The first order diffracted beams emerging from the
two AOMs were passed through two quarter-wave plates ($Q1, Q2$), to 
render one 
beam right circularly polarised (R) and the other left-circularly
polarised (L). The two beams were then combined at a non-polarising beam 
splitter 
(BS2). At each exit face of the final beam splitter, a polariser ($P3, 
P4$)
was kept, followed by a detector ($D3, D4$). The undiffracted beams 
terminated in beam dumps (BD) as shown in Fig.~\ref{fig2.schematic}. 
By keeping one of the polarisers (e.g.,  $P3$ ) fixed and changing the  orientation of the other polariser ($P4$), the relative angle between the polarisers,   $\phi_{34}$ could be 
varied continuously from $0 ^\circ$ to $360^\circ$. This results in a corresponding variation in the geometric phase. As we show below, this also causes a modulation of the intensity cross-correlation of the light reaching the two detectors. In order to measure this, 
the intensities of light reaching
detectors $D3$ and $D4$, for each orientation of $P4$,  was recorded  for a certain length of time  on a digital storage oscilloscope 
and cross correlated offline.   { Thus, from the recorded time series, one may determine first and second order correlations. }

In striking contrast to the other interference 
experiments which require a source of coherent 
light, the two-photon intensity-interferometery   needs  
incoherent sources as \hbt correlations  
vanish for a  laser light source. 
One could use a thermal source such as a mercury vapour lamp. 
However, the coherence time of natural  thermal light sources is too 
short for bunching effects to be discernible by present day 
solid-state photo-detectors.  Most two-photon interferometry experiments simulate  thermal light  by passing  laser light through a rotating ground-glass plate.  This introduces, random intensity and phase fluctuations  that occur  at a time scale detectable by present day solid-state detectors.   We  employed a different technique that utilises acousto-optic interaction. 

In an AOM, a radio-frequency (RF) electrical signal is applied to create a  travelling acoustic-wave grating that can
diffract light.  The frequency, intensity, and phase of light can be 
manipulated through acousto-optic 
interaction by suitably tailoring the RF  input to the AOM. In this experiment, however, only phase fluctuations were introduced.
As demonstrated recently~\cite{deepak},  an incoherent source may be created
by electronically introducing random phase jumps to the RF and 
imprinting these onto laser light.   
The phase 
evolves  undisturbed for a time $T$,  and then is changed
by a random amount $\delta$.
The phase jump $\delta$ 
is uniformly randomly distributed over the circle and  $T$  
has the distribution of a truncated exponential:
\begin{eqnarray}
\label{eq.1} P (T) &=& \frac{1}{T_c} e^{-\frac{T}{T_c}}~, 
\end{eqnarray}
 where $1 ~\mu s < T < 100~\mu s$, and $T_c$ = $10~\mu s$.  $T_c$ represents 
 the timescale over which the coherence of the light 
beam is lost.
Such a distribution (in untruncated form) has been discussed~\cite{loudon}
in connection with the emission of a single atom interrupted by 
collisions. Each AOM in the setup was independently  phase modulated in this manner  to derive two
independent phase-incoherent sources (S1 and S2 in Fig. ~\ref{fig2.schematic}) from the same laser light. 

{\it Theory:}  {We now examine  $E_3$ and $E_4$,  the  light fields reaching detectors D3 and D4,  in terms of   $E_1$ and $E_2$, the light fields  emerging from Q1 and Q2. Using the helicity basis vectors 
$\ket{R} = \left(\begin{array}{c}1\\ 
                              0\end{array}\right)$  and $\ket{L} = \left(\begin{array}{c}0\\ 
                              1\end{array}\right) $
we may represent 
 linear states by 
$\ket{\varphi_i}=\frac{1}{\sqrt{2}}
(e^{-i\varphi_i}\ket{R}+e^{i\varphi_i}\ket{L}$). 
We introduce projection matrices $P_K = \vert K\rangle\langle K \vert$ onto polarisation states $\ket{K}$, where K= R, L, 3, 4. More explicitly, 
\begin{equation}
P_{R} = \left(\begin{array}{c}1\;\;0\\ 
                              0\;\;0\end{array}\right);\;P_{L} = 
\left(\begin{array}{c}0\;\;0\\ 
                              0\;\;1\end{array}\right); \;  
P_{i} = {\frac{1}{2}}
\left(\begin{array}{lr}1\;\;\;\; e^{-i2\varphi_{i}}\\
e^{i2\varphi_{i}}\:\;1\end{array}\right)
\end{equation} 
Representing polarisation vectors with  Greek  incides and using the Einstein summation convention (for the Greek, but not for the Latin indices), }we can write
\begin{eqnarray}
{{E_i^\alpha}} &=& 
\frac{1}{\sqrt{2}}{{P}}_i^{\alpha\beta}~[\,\epsilon_i{{P}}_{ 
L}^{\beta\gamma}
\, {{E_2^{\gamma} }}\, u_{i2} \,+ \, {{P}}_{R}^{\beta\gamma} \,
{{E_1^{\gamma}}}\, u_{i1}\,] \nonumber\\
 {{{\bar {E}}_i^{\alpha}}} &=& \frac{1}{\sqrt{2}}[\,{\bar
u}_{i2}\, {{{\bar{E}}_2^{\gamma}}}\, {{P}}_{ L}^{\gamma 
\beta}\epsilon_i 
\,+\,
{\bar u}_{i1}\, {{{\bar{E}}_1^{\gamma}}}\, {{P}}_{R}^{\gamma\beta}\,]
~ {{P}}_i^{\beta\alpha}~,
\label{eq:eq2}
\end{eqnarray}
 where the overbar stands for complex
conjugation; $\epsilon_i$ is 
a pure phase $\epsilon_3=1$, $\epsilon_4=-1$ due to  the  geometry of the Mach-Zehnder
setup. 
  {The 
quantities of interest for the AB effect are  the different correlations at detectors D3 and D4. The 
first order correlations or intensities are
\begin{equation}
{\cal I}^1_{i}(\tau = 0) = \expect{I_i }=\expect{{\bar E}_i^\alpha E_i^\alpha}; \;\;\; i = 3,4
\label{first}
\end{equation}
Note that each detector $i$ has contributions from both $E_1$ and $E_2$. In our experiment, the light intensities were adjusted so that  $\expect{\bar{E_1} E_1} = \expect{\bar{E_2} E_2}, {\it i.e., }  \expect{I_1} = \expect{I_2}$. }
 The second order cross correlations are
\begin{equation}
{\cal G}^2_{34}(\tau) = \frac{\expect{I_3(0) \, I_4(\tau)}} {\expect{I_3}\expect{I_4}}=\frac{\expect{{\bar E}^\alpha_3(0) E^\alpha_3(0) {\bar E}^\beta_4(\tau)  E^\beta_4(\tau)}}{{\expect{{\bar E}^\alpha_3E^\alpha_3}}{\expect{{\bar E}^\beta_4 
 E^\beta_4}}}
\label{cross}
\end{equation}
and self correlations
\begin{equation}
{\cal G}^2_{ii}(\tau) = \frac{\expect{I_i(0) \, I_i(\tau)}} {\expect{I_i}^2}=\frac{\expect{{\bar E}^\alpha_i(0) E^\alpha_i(0) {\bar E}^\beta_i(\tau)  E^\beta_i(\tau)}}{{\expect{{\bar E}^\alpha_iE^\alpha_i}}^2}
\label{self}
\end{equation}
where $I_3 (0)$ and $I_4 (\tau)$ are the intensities measured at $D3$ 
at time $0$ and at $D4$ at time  $\tau$ respectively and the average $\langle  \ldots \rangle$
is a time average over the integration time $T_{{int}}$,  for a given setting of the polarisers P3 and P4 :
\begin{eqnarray} && \expect{f} = \frac{1}{T_{int}} \int_0^{T_{{int}}}~ {f(t) dt}
\label{def} 
\end{eqnarray}

 {Substituting in Eqs. \ref{cross} and Eq. \ref{self}  from Eq. \ref{eq:eq2} and multiplying the terms explicitly yields 16 terms, of which 10 vanish on time averaging.  Use is made of the fact that the $2\times2$ Hermitean
projection matrices ${{P}}$ satisfy ${{P}}^2={P}$. Two of the surviving terms have a sequence of projections  of the form $P_R P_3 P_L P_4 P_R$, which define a closed trajectory on the \poi sphere.} This gives rise to the geometric phase $\Phi_G = \Omega /2$, that is,   half the solid angle enclosed.  Simple algebra yields, for the cross correlation 
\begin{eqnarray}
{\cal G}^2_{34}(\tau = 0) &=&  1-\frac{1}{2}
\cos
\left(\phi_D+\frac{\Omega}{2}\right)~,
\label{corr1} \end{eqnarray}. 

and for the self correlation 
\begin{eqnarray}
{\cal G}^2_{ii}(\tau  = 0) &=&  1 + \frac{1}{2} 
\cos
\left(\phi_D\right)~, 
\label{corr2} \end{eqnarray}
where $\phi_D$ represents the dynamical phase 
(which, in our Mach-Zehnder interferometer has been  adjusted to zero) 
and $\frac{\Omega}{2}$
is the geometric phase. The negative sign for the second term 
(in Eq.~\ref{corr1})
is due to the Mach-Zehnder configuration. 

 {On similar lines,  using Eq. \ref{eq:eq2} in Eq. \ref{first}, the first order correlation may be  evaluated. Of the four terms in the product, only two survive, with products of the form $P_i P_R P_i,  (i = 3,4)$,  which depend only on the intensities of the sources and not the geometric phase. Thus, 

\begin{equation} 
  {{\cal I}_{i} ^1} = {\frac {\langle I_1 + I_2\rangle}{4}},  \;\;\;\;\;\; i = 3,4
\label{g1}
\end{equation}}

{ {The  geometric phase appears only in the second order intensity cross correlation (Eq.~\ref{corr1}), and is absent both in the 
 second order intensity self correlation  (Eq.~\ref{corr2})
and  the first order correlation (Eq. \ref{g1}). Thus, this effect is analogous to the two-particle non-local AB effect. }

\begin{figure}
\onefigure [scale=0.32]{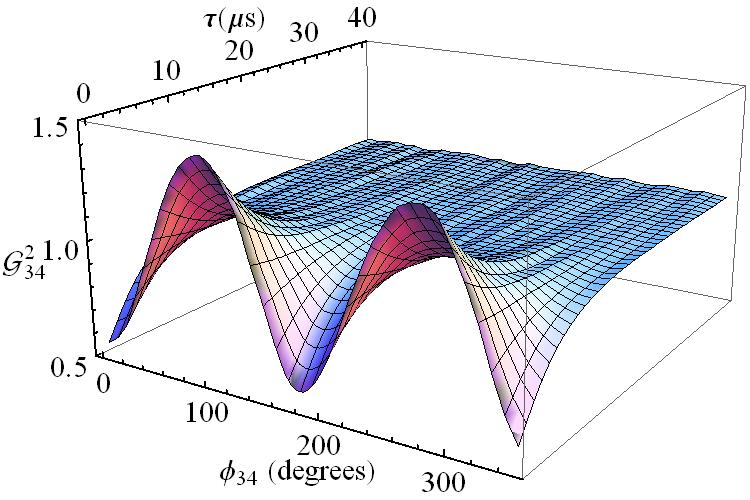}
\caption{(Colour online) Figure shows the experimentally determined  cross correlation, ${\cal G}^2_{34}$,
as a function of
$\phi_{34}$ the relative angle between the polarisers $P3$ and $P4$,  and as function of  time delay $\tau$. Clearly,  the cross correlation ${\cal 
G}^2_{34}$ is 
modulated by the 
geometric phase. }
\label{fig3.measurement}
\end{figure}

{\it Experimental Results :} 
The main results of our experiment are displayed in the three-dimensional 
plot of Fig.~\ref{fig3.measurement},
which shows the quantity of interest, ${\cal G}^2_{34}$, as a function
of the time delay, $\tau$, and the relative angle, $\phi_{34}$, between the 
linear polarisers $P3$ and $P4$ in front of the two detectors $D3$ and 
$D4$ respectively.   {In the experiment $P3$ was kept fixed and $P4$ was rotated. In effect,   $\phi_{34}$ is a measure of the geometric phase.} 
\begin{figure}
\onefigure[scale=0.32]{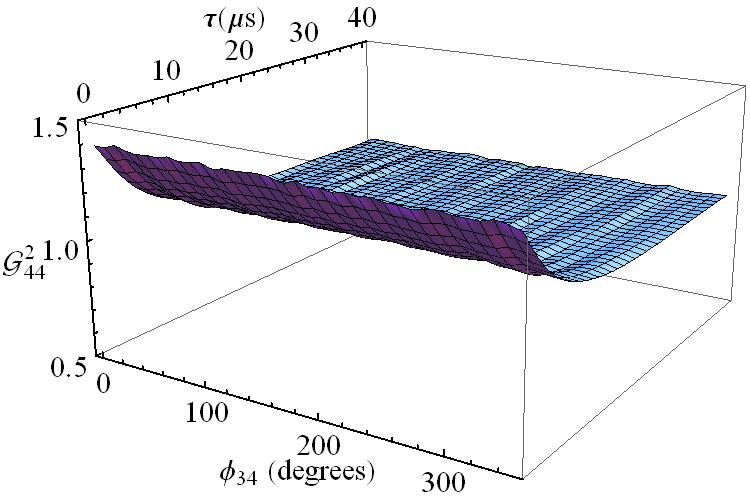}
\caption{(Colour online) Figure shows the self correlation ${\cal G}^2_{44}$ 
as 
a function of
$\phi_{34}$ the relative angle between the polarisation angles of the 
two detectors, and of the time delay $\tau$. 
Note that the self correlation (${\cal 
G}^2_{44}$ is plotted) is independent of $\phi_{34}$ . 
 {$\phi_{34}$ is varied by turning the polaroid P4 in the 
observed channel.}
Thus the geometric 
phase effect of Fig.~\ref{fig3.measurement} is a
purely nonlocal effect.} 
\label{fig4.nonlocal}
\end{figure}
\begin{figure}
\onefigure[scale=0.45]{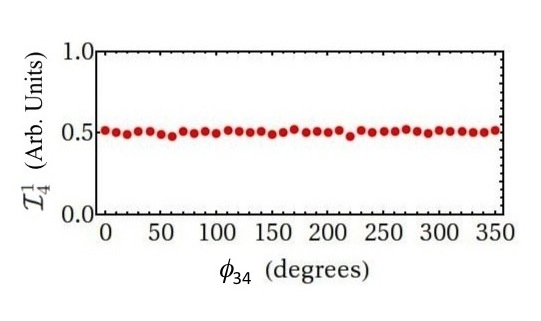}
\caption{ {(Colour online) 
${\cal I}^1_4$ as a function of $\phi_{34}$ as obtained from the experiment. $\phi_{34} = \phi_3 - \phi_4$ is varied by turning the polaroid P4 in the 
observed channel while keeping P3 fixed.}} 
\label{fig5.intensity}
\end{figure}

 For zero delay ($\tau = 0$)  between the two detectors, the cross 
correlation ${\cal G}^2_{34}$ was found to vary sinusoidally from $\sim0.5$ 
to $\sim1.5$ as $\phi_{34}$ is varied.  Clearly, this is 
due to the geometric phase $\Phi_G =\Omega/2=2\phi_{34}$. 
For larger values of  $\tau$  the amplitude of the sinusoidal variation is found to be progressively diminished, 
 till 
finally, for $\tau >T_C$, ${\cal G}^2_{34}$ remains nearly 
constant 
at 
$1$, {\it i.e., } correlations that are maximum for zero delay gradually decrease, and are absent for  delays larger than the coherence time. 
At the regions of constructive intereference (e.g. $\phi_{34} = 
90^\circ$), 
the value of $  {\cal G}^2_{34}$ starts from $ ~1.5$ for $\tau = 0$ as is 
expected for a source with pure phase fluctuations~\cite{baym}.
Note that in regions of destructive interference, 
the cross correlation starts from $0.5$, a value lower than
that for a coherent source. This has been discussed in various contexts 
earlier~\cite{hpaul}.

In order to illustrate that it is a purely nonlocal effect we also 
plot in Fig.~\ref{fig4.nonlocal}, the second order self
correlation.  In contrast to the cross-correlation, the self-correlation  remains 
unaltered as the relative angle between the polarisers is
varied. This may be easily understood by visualising the 
trajectory on the  \poi sphere.     {For example, for ${\cal G}^2_{44}$ the trajectory  is obtained by the sequence $P_R P_4 P_L P_4 P_R$  and defines an arc connecting  the three
points  $R,4,L$   and does not enclose any solid angle; 
hence the absence of 
a geometric phase~\cite{panch}.

Having established the non-local nature of this effect , we next verify its absence in the lower order correlation.  Using  the same experimental data that was used above, we now examine the quantity ${\cal I}^1_4$. }   {In Fig.~\ref{fig5.intensity} we display the quantity ${\cal I}^1_{i}(\tau = 0)$ as a function
of the setting $\phi_4$ of the polaroid P4 in front of detector D4. Note that this curve is 
practically constant, showing that turning the polaroid P4 does not
affect the lower order correlation function, that is, the geometric phase 
is not seen in the lower order correlations.}\\
{\it Discussion:} 
 The  experiment and  the theoretical analysis  
presented here are  purely classical, in contrast to the earlier quantum mechanical treatment  
\cite{Mehta} that predicted similar results, and  the recent photon-counting  intensity interferometry experiments\cite{martin} that verified it. 
 {In this Letter we develop the analogy between {\it 
classical} polarised light and the quantum nonlocal AB effect.  Further, we explicitly show its non-local nature, and its absence in lower order correlations.}
 
The classical approach  has the advantage that the physical ideas are
easy to grasp. In fact, Hanbury Brown and Twiss were originally motivated by
their classical experience with radio waves to propose
the corresponding optical experiment. The quantum interpretation 
of this classically simple experiment led to
profound changes in our understanding of quantum optics and coherence.
          
In addition to being completely in the classical domain, our 
experiment has another novelty   {- the use of a source that has phase fluctuations only. }
In the case of thermal or pseudothermal light ({\it i.e., } one with both phase and intensity fluctuations), ${\cal G}^2$ 
varies between $1$ and $2$. In contrast, for pure phase modulation,
one expects a variation from $0.5$ to $1.5$. 
Baym, in his review 
article ~\cite{baym} that examines the \hbt effect in a wide variety of physical contexts ranging
from nuclear physics to astronomy, had reached this conclusion in his discussion  of  the case of pure phase fluctuations. However, this has hitherto not been verified, as pure "phase-only" fluctuating sources  have not been available. In our experiment, such a source has  been realised by  tailoring fluctuations in light by the acousto-optic technique, and the result of Baym verified.   

 Two salient features of the geometric phase measured in our experiment 
are its purely non-local nature and its absence in lower order 
 {correlations},  in contrast to earlier work on geometric phase 
and intensity interferometry \cite{brendel}. 
 These specific features give rise to the possibility of the application of this nonlocal AB 
effect 
as a two-photon deflector. We refer again to Fig.~\ref{fig2.schematic},  where we now replace the polariser $P4$ by a  ``graded polariser" - one where the axis of polarisation changes in orientation gradually as one moves across the polariser.  In intensity interferometry this would  introduce a geometric phase
gradient. 
As is well known, a phase gradient in a wave is
equivalent to a deflection or a change of wave vector. In
the present experiment, the phase change appears {\it only}
in the intensity cross correlations. Thus, we may 
generate a ``correlated-photon deflector''  which  affects the intensity-intensity correlation, leaving 
individual  photons
unaffected. 
Such a device may have applications in creating and manipulating correlated photons.

 We close with a few classical remarks
which may have interesting quantum interpretations.
Let us note that the geometric phase is unbounded.
As one turns the polariser  $P4$, the geometric phase keeps
accumulating without bound and the visibility of the  interference pattern remains undiminished. This is in sharp contrast to the dynamical phase: as one increases the dynamical phase, by separating the detectors, the interference pattern is washed out
when their spatial (temporal) separation exceeds the coherence length (time).  This difference is due to the fact
that the geometric phase is achromatic and affects all
frequencies equally. Thus the geometric phase effects are
less susceptible to decoherence than the corresponding
dynamical effects.
This illustrates a 
point often made in
the quantum computation literature that geometric and
topological effects are robust against decoherence. This
is crucial to the development of quantum computers.

At the level of classical information theory, our experiment can be 
interpreted as a way of delocalising information. 
Consider the experimental setup above and make the following minor changes:
We choose $P3$ and $P4$ to be orthogonal linear polarisations 
represented 
by antipodal points on the equator of the 
\poi sphere.
We take the sources to be elliptically polarised and lying on the 
great circle orthogonal to the line joining the  antipodal points $3$ 
and $4$.
If the angle $\phi$ between the elliptical polarisations of the two sources 
is varied in time $\phi(t)$, one would find that
there is a corresponding modulation in the cross-correlations of the intensities detected at $D3$ and $D4$.  
The signal $\phi(t)$ can be viewed as carrying information. 
However, the two beams emerging from the experiment are of fixed linear polarisation, 
and each beam by itself appears thermal in its self 
correlation.
It is only by looking at their cross correlations that one
can recover the originally impressed signal $\phi(t)$. Thus, in this example,
information is stored in a completely delocalised manner. 
In quantum information theory, one
expects that it would be similarly possible to have 
apparently thermal beams carrying information 
entirely in their quantum cross correlations.

To summarise, we have  demonstrated the existence of a non-local geometric phase in  the intensity-intensity correlations of classical incoherent light, that is
not seen in the lower order correlations. 
This two-photon Pancharatnam phase is the classical optical
analogue of the multi-particle AB effect \cite{heiblum}.  
As the trajectory
of light over the \poi sphere introduces a phase shift observable
only in the intensity-intensity correlation, it provides a means of
deflecting  the two-photon wavefront, while having no effect on single photons.  The experiments were performed using sources that had
 pure phase fluctuations.  We expect that the results presented here 
will be of  interest  for applications in the realm of classical and quantum communication and cryptography.

\acknowledgments

\bibliographystyle{eplbib}
\bibliography{referenceshbt}

\begin{thebibliography}{10}
\expandafter\ifx\csname url\endcsname\relax\def\url#1{\texttt{#1}}\fi

\bibitem{Strekalov}
\Name{Strekalov D.~V., Stowe M.~C., Chekova M.~V. \and Dowling J.~P.}
  \REVIEW{J. Mod. Opt. }{49}{2002}{2349}.

\bibitem{Saleh}
\Name{Saleh B.~E., Abouraddy A.~F., Sergienko A.~V. \and Teich M.~C.}
  \REVIEW{Phys. Rev. A }{62}{2000}{043816}.

\bibitem{Pittmann}
\Name{Pittmann T., Shih Y., Strekalov D. \and A.V. S.} \REVIEW{Phys. Rev. A
  }{52}{1995}{R3429}.

\bibitem{Bouwmeester}
\Name{Eisenberg H., Hodelin J., Khoury G. \and Bouwmeester D.} \REVIEW{Phys.
  Rev. Lett. }{94}{2005}{090502}.

\bibitem{shihbook}
\Name{Shih Y.} \Book{An Introduction to Quantum Optics : Photon and Biphoton
  Physics} (CRC Press (U.S.)) 2011.

\bibitem{Bennick-Boyd}
\Name{Bennick, R. S., Bentley, S. J. \and Boyd, R. W.} \REVIEW{Phys. Rev. Lett.
  }{89}{2002}{113601}.

\bibitem{Abouraddy}
\Name{Abouraddy A., Saleh B., Sergienko A. \and Teich M.} \REVIEW{Phys. Rev.
  Lett. }{87}{2001}{123602}.

\bibitem{Chapter-21}
\Name{Saleh B.~E. \and Teich M.~C.} \Book{Noise in Classical and Quantum
  Photon-Correlation in Advances in Information Optics and Photonics ed. A.T.
  Frieberg and R. Dandiker in} (SPIE Press, Bellingham, WA) 2008.

\bibitem{unified-theory}
\Name{Erkmen B.~I. \and Shapiro J.~H.} \REVIEW{Adv. in Optics and Photonics
  }{2}{2010}{405}.

\bibitem{Klyshko}
\Name{Klyshko D.} \REVIEW{Phys. Lett. A }{146}{1990}{93}.

\bibitem{Burlakov}
\Name{Burlakov, A. V., Chekova, M. V., Mamaeva, Yu B., Karabutova, O. A.,
  Korystov, D. Y. \and Kulik, S. P.} \REVIEW{Laser Phys. }{12}{2002}{1}.

\bibitem{Shih-review}
\Name{Shih Y.} \REVIEW{Rep. Prog. Phys. }{66}{2003}{1009}.

\bibitem{EPL_Scarcelli-Shih}
\Name{Scarcelli G., Valencia A. \and Shih Y.} \REVIEW{EPL }{68}{2004}{618}.

\bibitem{Valencia}
\Name{Valencia A., Scarcelli G., D'Angelo M. \and Shih Y.} \REVIEW{Phys. Rev.
  Lett. }{94}{2005}{063601}.

\bibitem{Friberg}
\Name{Shirai T., Setala T. \and Frieberg, A. T.} \REVIEW{Phys. Rev. A
  }{84}{2011}{041801}.

\bibitem{Zhou}
\Name{Zhou Y., Simon J., Liu J. \and Shih Y.} \REVIEW{Phys. Rev. A
  }{81}{2010}{043831}.

\bibitem{Karmakar}
\Name{Karmakar S., Zhai Y., Chen H. \and Shih Y.} \Book{The first ghost image
  using sun as light source} in proc. of \Book{Quantum Electronics and Laser
  Science Conference.} (Optical Society of America) 2011 p. QFD3.

\bibitem{Dheera}
\Name{Venkataraman D., Hardy N., Wong~Francis, N. C. \and Shapiro, Jeffrey H.}
  \REVIEW{Opt. Lett. }{36}{2011}{3684}.

\bibitem{Mehta}
\Name{Mehta P., Samuel J. \and Sinha S.} \REVIEW{Phys. Rev. A
  }{82}{2010}{034102}.

\bibitem{panch}
\Name{Pancharatnam S.} \REVIEW{Proc. Ind. Acad. Sci. A}{44}{1956}{247}.

\bibitem{punch}
\Name{Ben-Aryeh Y.} \REVIEW{Journal of Optics B: Quantum and Semiclassical
  Optics }{6}{2004}{R1}.

\bibitem{samuelsson}
\Name{Samuelsson P., Sukhorukov E.~V. \and B\"uttiker M.} \REVIEW{Phys. Rev.
  Lett. }{92}{2004}{026805}.

\bibitem{ab}
\Name{Aharonov Y. \and Bohm D.} \REVIEW{Phys. Rev. }{115}{1959}{485}.

\bibitem{SamCS}
\Name{Samuel J.} \REVIEW{Curr. Sc. }{66}{1994}{781}.

\bibitem{PT}
\Name{Batelaan A. \and Tonomura A.} \REVIEW{Phys. Today }{62}{2008}{38}.

\bibitem{wuyang}
\Name{Wu T.~T. \and Yang C.~N.} \REVIEW{Phys. Rev. D }{12}{1975}{3845}.

\bibitem{Sam-Hema}
\Name{Hariharan P., Ramachandran H., Suresh K. \and Samuel J.} \REVIEW{J.
  Modern Optics }{44}{1997}{707}.

\bibitem{heiblum}
\Name{Neder I., Ofek N., Chung Y., Heiblum M., Mahalu D. \and Umansky V.}
  \REVIEW{Nature }{448}{2007}{333}.

\bibitem{deepak}
\Name{Pandey D., Satapathy N., Meena M.~S. \and Ramachandran H.} \REVIEW{Phys.
  Rev. A }{84}{2011}{042322}.

\bibitem{loudon}
\Name{Loudon R.} \Book{The Quantum Theory of Light} (Oxford University Press)
  2010.

\bibitem{baym}
\Name{Baym G.} \REVIEW{Acta. Phys. Polonica B}{29}{1998}{1839}.

\bibitem{hpaul}
\Name{Paul H.} \REVIEW{Reviews of Modern Physics }{54}{1982}{1061}.

\bibitem{martin}
\Name{Martin A., Alibart O., Flesch J.-C., Samuel J., Sinha S., Tanzilli S.
  \and Kastberg A.} \REVIEW{EPL (Europhysics Letters) }{97}{2012}{10003}.

\bibitem{brendel}
\Name{Brendel J., Dultz W. \and Martienssen W.} \REVIEW{Phys. Rev. A
  }{52}{1995}{2551}.

\end{thebibliography}

\end{document}